\documentclass[floatfix, noshowpacs,  twocolumn, amsmath, amssymb, aps, prl, superscriptaddress]{revtex4-1}

\usepackage{graphicx, xcolor}
\usepackage[caption=false]{subfig}
\usepackage{soul}
\usepackage{amsmath}
\usepackage{adjustbox}
\usepackage{tikz}
\usepackage{hyperref, stackengine, tabularx, xcolor, graphicx}
\usepackage{footmisc}
\usepackage{ulem}

\newcommand{\ket}[1]{\left| #1 \right>}

\newcommand{\agrave}{\`a}

\newcommand{\subfigimg}[3][,]{%
	\setbox1=\hbox{\includegraphics[#1]{#3}}
	\leavevmode\rlap{\usebox1}
	\rlap{\hspace*{210pt}\raisebox{\dimexpr\ht1-2\baselineskip}{#2}}
	\phantom{\usebox1}
}

\begin{document}

\title{Nonlinear Inverse Spin Galvanic Effect in Anisotropic  Disorder-free Systems}

\author{Iryna Miatka}
\affiliation{Dipartimento di Matematica e Fisica, Universit{\agrave} degli Studi Roma Tre,
	Via della Vasca Navale 84, 00146 Rome, Italy}

\author{Marco Barbieri}
\affiliation{Dipartimento di Scienze, Universit{\agrave} degli Studi Roma Tre,
	Via della Vasca Navale 84, 00146 Rome, Italy}

\author{Roberto Raimondi}
\affiliation{Dipartimento di Matematica e Fisica, Universit{\agrave} degli Studi Roma Tre,
	Via della Vasca Navale 84, 00146 Rome, Italy}

\begin{abstract}	

Spin transport phenomena in solid materials suffer limitations from spin relaxation associated to disorder or lack of translational invariance. Ultracold atoms, free of that disorder, can provide a platform to observe phenomena beyond the usual two-dimensional electron gas. By generalizing the approach used for isotropic two-dimensional electron gases, we theoretically investigate the inverse spin galvanic effect in the two-level atomic system in the presence of anisotropic Rashba-Dresselhaus spin-orbit couplings (SOC) and external magnetic field. We show that the combination of the SOC results in an asymmetric case: the total spin polarization considered for a small momentum has a longer spin state than in a two-dimensional electron gas when the SOC field prevails over the external electric field. Our results can be relevant for advancing experimental and theoretical investigations in spin dynamics as a basic approach for studying spin state control.

\end{abstract}
\maketitle

\section{Introduction} 
The spin galvanic effect (SGE), i.e  
spin polarization-charge current interconversion, is currently the focus of intensive theoretical and applied research~\cite{Ganichev2016}. Quite generally, in the absence of inversion symmetry, polar and axial vectors may transform similarly and hence a coupling between the spin polarization and the charge current density can occur~\cite{Ivchenko1978}. In solid-state systems, as in semiconductors and metals, the key microscopic mechanism  is provided by the spin-orbit coupling (SOC)~\cite{Edelstein90,Aronov89}.  Both the Rashba and the Dresselhaus spin-orbit coupling  (RSOC and DSOC, respectively), have been considered. RSOC arises as a result of the structure inversion asymmetry (SIA) in a semiconductor quantum well or at a metal interface.
RSOC was first proposed for non-centrosymmetric wurtzite semiconductors~\cite{Rashba} and, later, also for 2D electron gases~\cite{bychkov1984properties}. Besides, spin-dependent splitting in heterojunctions has provided the basis for the spin transistors~\cite{DattaDas}. RSOC may be seen as arising from the   Zeeman interaction between the magnetic moment component parallel to the spin and the effective magnetic field moving with the electron~\cite{Newperspectives}. Recently, RSOC was detected in new materials such as metal surfaces, bulk and interfaces materials of semiconductors~\cite{1,2,3,4,5,13} and heavy metals~\cite{3}. Moreover, RSOC plays an important role even in more exotic areas like
the emergence of topological states in insulators and Majorana fermions in topological superconductors~\cite{Hasan,Beenakker}. 
DSOC, on the other hand, arises in the bulk as a consequence of the bulk inversion asymmetry (BIA)~\cite{Dresselhaus55}. 

The  inverse SGE manifests as a non-equilibrium spin polarization due to an applied electric field. In solid-state systems, dissipation due to disorder created by impurities and imperfections leads to steady-state conditions with a DC charge current proportional to the applied electric field. In this linear-response regime, both the charge current and the non-equilibrium spin polarization are proportional to the electric field~\cite{Edelstein90,rai2014microscopic}. In such a situation the electron drift velocity remains much smaller than the Fermi velocity, and the spin polarization also remains small. Recently, Vignale and Tokatly~\cite{vignale2016theory} have considered the case of a perfectly clean two-dimensional electron gas (2DEG) with RSOC and have provided an elegant exact solution of the associated model, which allows to analyze the SGE and its inverse in the nonlinear regime, where electrons can be accelerated to high velocities.

While the discussion on disorder-free systems might appear purely academic, there exists the possibility of investigating spin-orbit coupling effects in cold atoms~\cite{114}, that are free from strong disorder. However, the synthetic fields produced on atoms by the action of Raman fields~ \cite{zhu2006spin,Liu09,lin2011spin, su2016rashba} result in an anisotropic coupling, which can be seen as a combination of RSOC and DSOC.

In this paper we extend the theory of Vignale and Tokatly of nonlinear effects in inverse SGE to anisotropic spin-orbit couplings in the presence of a Zeeman field. We demonstrate that adding this complexity still allows for an analytical solution of the model, and its phenomenology can be fully analyzed.  In  particular, we investigate the dynamics of the spins in the adiabatic regime with Rashba spin-orbit coupling. To investigate the spin-dependent evolution we directly solve the Schr\"odinger equation in terms of spinors and present the analytical solutions. In addition, we describe the average non-equilibrium spin polarization and tunability of the system for different values of the two Rashba coefficients. We then compare the result with the adiabatically-approximated spin polarization.

\section{ The model  and its solution}  
We consider a collection of spin-one half atoms, which is driven by a constant and homogeneous electric field defined,  in a vector gauge, from the vector potential ${\bf A}=-E\,t\,\hat{\textbf{x}}$. 
Due to the  translational invariance of the system, the system single-particle eigenstates are plane waves with a given momentum. 
The system evolution is then governed by the Hamiltonian:
\begin{equation}
\begin{aligned}
H({\bf p},t)= H_0+H_{SO}+H_{E}+H_{ext},
\end{aligned}
\label{3}
\end{equation}
\noindent where $H_0= \frac{1}{2m}[{\bf p}+e{\bf A}]^2$ corresponds to the kinetic part and $H_{SO}$ is Rashba spin-orbit interacting term: $H_{SO}= \alpha_1\,\sigma_x\,p_y-\alpha_2\,\sigma_y\,p_x$,
$p_{x,y}$  being the in-plane  components of the momentum operator and $\sigma_{x,y}$  the Pauli matrices  associated to the spin degree of freedom~\cite{note1}. {\color{black} This coupling produces the same effects as a magenetic field, which we will call Rashba field. In the standard treatment of RSOC and DSOC, one would have instead Hamiltonians of the kind $H_{RSOC}= \alpha\,(\sigma_x\,p_y-\sigma_y\,p_x)$, and $H_{DSOC}= \beta\,(\sigma_x\,p_y+\sigma_y\,p_x)$, respectively. The asymmetric case we are treating can then be considered as a combination of the two effects with strengths $\alpha=(\alpha_1+\alpha_2)/2$ and $\beta=(\alpha_1-\alpha_2)/2$. When the electric field is applied, minimal coupling results in an Edelstein term $H_E=\alpha_2eEt \,\sigma_y$, which amounts to the presence of an effective field growing with time. In addition, we also include a Zeeman coupling with an external magnetic field $H_{ext}=-\mu_B{\bf B}\cdot{\boldsymbol{\sigma}}$. 
The occurrence of such coupling terms is not uniquely attributed to the application of actual fields and the simultaneous presence of RSOC and DSOC mechanisms: these may be effective descriptions. For cold atom systems, the Hamiltonian (\ref{3}) is derived from a single-particle Hamiltonian where two hyperfine states of the ground level are coupled via Raman lasers, whose phase and intensity depend on the position. The projection of the Hamiltonian onto the manifold of these two states generates the non-Abelian gauge potential~\cite{zhu2006spin,Liu09}. The rotation of the spin axes in this gauge potential reveals asymmetric spin-orbit coupling, which can be viewed as a combination of Rashba and Dresselhaus spin-orbit couplings.}



\begin{figure}
	\includegraphics[width=0.9\linewidth]{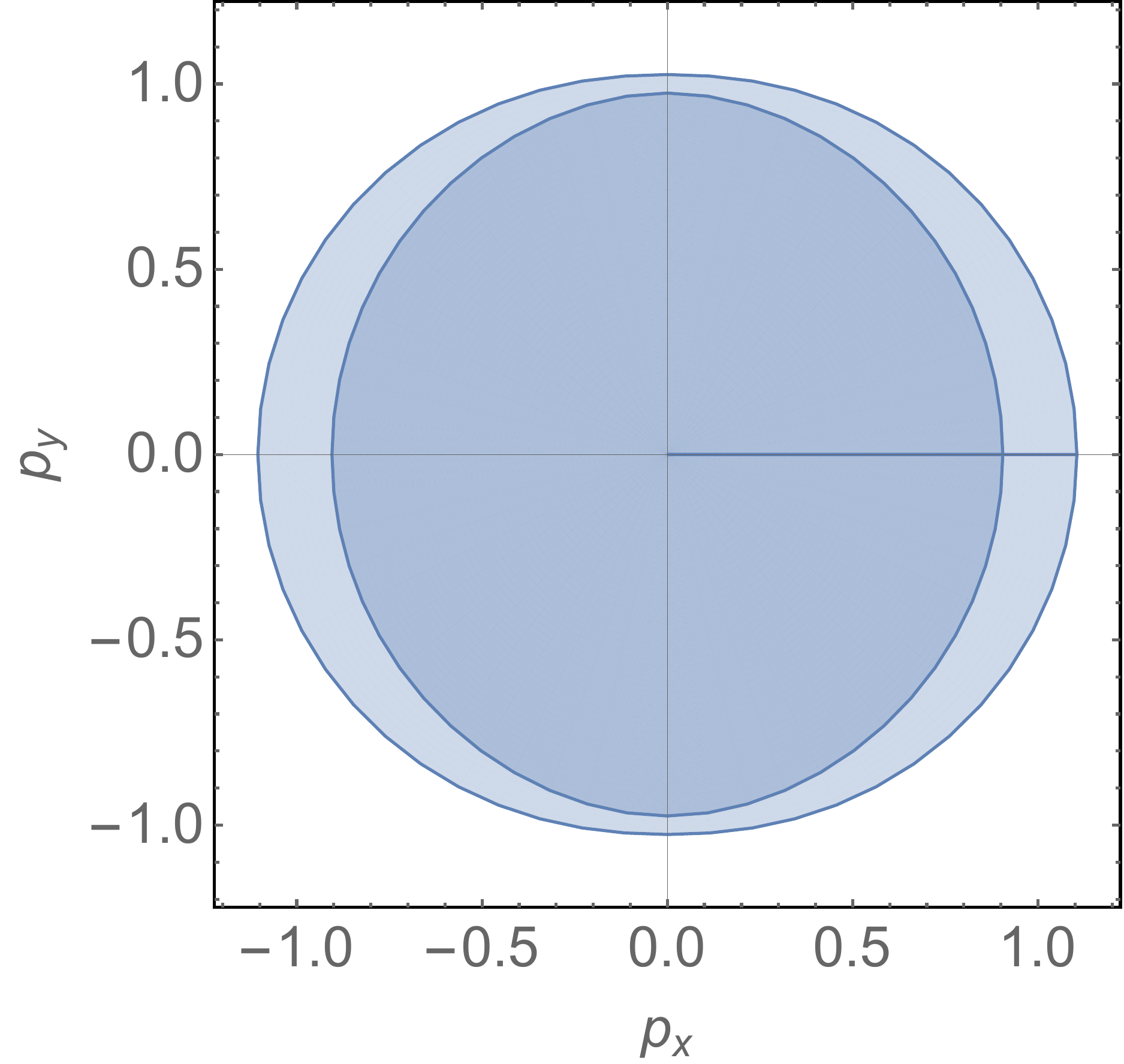}
	\caption{\textbf{Fermi surfaces with  anisotropic SOC.} The Fermi surfaces associated to the two lower momentum $p_2(p_F,\theta)$, and to the higher  momentum $p_1(p_F,\theta)$ are shown as a function of the momentum components (in units of $p_F$). Here we set $m\alpha_2=0.1p_F$, and $\alpha_1/\alpha_2=1/4$. States below $p_2(p_F,\theta)$ are fully occupied (dark blue region), thus we only consider the contribution of states in the light blue region.}
	\label{ellipse}
\end{figure}

To begin with, we consider  the effect of having two different Rashba coefficients $\alpha_1$ and $\alpha_2$; in the absence of external fields, the eigenvalues of the spin-orbit part of the Hamiltonian read
\begin{equation}
\label{p}
E_{1,2}=\frac{p^2}{2m}\mp\sqrt{\alpha_1^2p_y^2+\alpha_2^2p_x^2}{\ }
\end{equation}
\noindent being $p_x= p\,\cos\theta$ and $p_y= p\,\sin\theta$. The different coefficients break the circular symmetry of the constant energy contour of the pure RSOC: the absolute value of the momentum now explictly depends on the angle $\theta$ between the $x$ axis and $\hat{\textbf{p}}$, giving an elongated shape to the contours:
 \begin{equation}
p_{1,2}(\theta, p_0)=\sqrt{m^2\alpha^2 (\theta)+p_0^2}\pm m \alpha (\theta)
\label{momentum}
\end{equation}
with $\alpha (\theta)=\sqrt{\alpha_1^2 \sin^2 (\theta)+\alpha_2^2 \cos^2 (\theta)}$ and $p_0^2/2m$ defining the energy of the contour and the Fermi momentum in the absence of SOC, Fig.~\ref{ellipse}.
Furthermore, the spin helicity of the eigenstates is no longer perpendicular to the momentum.

The implication of this asymmetry for the spin polarization dynamics can be assessed by directly solving the  Schr\"{o}dinger equation for the Hamiltonian in Eq.~({\ref{3}}). 
For a given Fermi momentum $p_0=p_F$, the energy levels with $p<p_2(\theta,p_F)$ have both spin states occupied and then do not contribute to the spin dynamics. Hence, 
we  consider only the states with $p_2(\theta, p_F)<p<p_1(\theta,p_F)$.

The interacting and free parts of the Hamiltonian ({\ref{3}}) can be written as a $2\times2$ matrix; following Ref.\cite{vignale2016theory} the y axis is  taken as the quantization direction for spin~\cite{note2}:
\begin{widetext}
\begin{equation*}
H=\begin{pmatrix}
-\alpha_2 \,p\,\cos\theta+\alpha_2\,eEt-\mu_B B_y & -\alpha_1\,p\,\sin\theta+\mu_B (B_x+iB_z)\\
-\alpha_1\,p\,\sin\theta+\mu_B (B_x-iB_z) & \alpha_2\,p\,\cos\theta-\alpha_2\,eEt+\mu_B B_y
\end{pmatrix}.
\end{equation*}
\end{widetext}

The above form is the one for the Landau-Zener problem\cite{Zener1932}, where the energy difference of two coupled levels varies linearly in time.  
The combination $\alpha_2\,eEt$ enters as an effective magnetic field due to the electric field. Following Ref.\cite{vignale2016theory} we refer to this term as the Edelstein field.
The two relevant energy scales are the kinetic energy $p^2/2m$ and {\color{black} the work $eE L_{SOC}$ performed by the electric field the over the spin-orbit length $L_{SOC}= \hbar (2m\alpha_2)^{-1}$. In the absence of relaxation, this is the most relevant length scale}.
In the following for the sake of simplicity we take units such that $\hbar =1$.
Notice  that in the present anisotropic SOC, we must consider the SOC length in the same direction as the applied field.  It is then useful to introduce an adiabaticity parameter $\gamma_p$ defined as the ratio of the above energy scales
\begin{equation}
\label{adiabaticity}
\gamma_p =\frac{ e E L_{SOC}}{p^2/2m}=\gamma_0\frac{p_0^2}{p^2},
\end{equation}
where $\gamma_0=\frac{eE}{\alpha_2 p_0^2}$.
The value of $\gamma_p$ controls how fast the spin system reacts to the electric field.
By resorting to the natural units of energy $\alpha_2 p \sqrt{\gamma_p}$, we introduce the following dimensionless quantities
\begin{equation}
\begin{aligned}
&\tau = {\alpha_2p\sqrt{\gamma_p}} \ t {\ \ \ \ \ \ \ \ \ \ \ }
\tau_p=\frac{\cos\theta}{\sqrt{\gamma_p}}\\
&\Delta_p=\frac{\sin\theta}{\sqrt{\gamma_p}}\frac{\alpha_1}{\alpha_2}{\ \ \ \ \ \ \ \ \ \ }
\xi_{x,y,z}=\frac{\mu_BB_{x,y,z}}{\alpha_2 \, p\, \sqrt{\gamma_p}}.
\end{aligned}
\label{5}
\end{equation}
 After the changes the Hamiltonian takes the form
\begin{equation}
H_p(\tau) = \begin{pmatrix}
\tau-\tau_p-\xi_y & -\Delta_p+\xi_x+i\xi_z \\
-\Delta_p+\xi_x-i\xi_z& -(\tau-\tau_p-\xi_y)
\end{pmatrix}.
\end{equation}


We then look for the solution of the time-dependent Schr\"{o}dinger equation as a spinor with the two amplitudes $U(\tau-\tau_p-\xi_y)$ and $V(\tau-\tau_p-\xi_y)$

\begin{equation}
\ket{\Psi(\tau)}=U(\tau-\tau_p-\xi_y)\ket \uparrow+V(\tau-\tau_p-\xi_y)\ket\downarrow
\label{7}
\end{equation}
in terms of which the Schr\"odinger equation becomes
\begin{equation}
\left\{
\begin{aligned}
i\dot U(\tau)&=\tau U(\tau)-\tilde \Delta_p\,V(\tau),\\
i\dot V(\tau)&=-\tau V(\tau)-\tilde\Delta_p^*\,U(\tau),
\end{aligned}
\right.
\label{SysEq}
\end{equation}
where $\tilde\Delta_p =\Delta_p-(\xi_x+i\xi_z)$.
By rescaling the time variable as $z=\sqrt{2}e^{-i\frac{3\pi}{4}}\tau$
and introducing the parameters $\nu_+=\frac{\tilde\Delta_p}{\sqrt{2}}e^{-i\frac{3\pi}{4}}$,   $\nu_-=\frac{\tilde\Delta_p^*}{\sqrt{2}}e^{i\frac{\pi}{4}}$,
and $\nu =\nu_+\nu_-=-\frac{i}{2}|\tilde\Delta_p|^2$
our system of equations becomes
\begin{equation}
\left\{
\begin{aligned}
\dot{U}(z)+\frac{z}{2}U(z)-\nu_+V(z)=0\\
\dot{V}(z)-\frac{z}{2}V(z)+\nu_-U(z)=0,
\end{aligned}
\right.
\end{equation}
whose solution can be found in terms of parabolic cylinder functions by writing
\begin{equation}
	\begin{aligned}
		&U(z)=D(\nu,z),\\
		&V(z)=\nu_-D(\nu-1,z).
	\end{aligned}
	\label{sol1}
\end{equation}

We emphasize that the above represents the extension to the anisotropic case of the
  solution originally found by Ref. \cite{vignale2016theory}. 
In the full normalized form,  two mutually orthogonal solutions read
\begin{equation}
	\begin{aligned}
		&U^{(1)}(\tau)=e^{-\pi |\tilde \Delta_p|^2/8}D(\nu,z),\\
		&V^{(1)}(\tau)=e^{-\pi |\tilde \Delta_p|^2/8}\nu_- D(\nu-1,z),\\
		&U^{(2)}(\tau)=-[V^{(1)}(\tau)]^*,\\
		&V^{(2)}(\tau)=[U^{(1)}(\tau)]^*.
	\end{aligned}
	\label{ind_sol}
\end{equation}
The chosen initial conditions, obtained by solving the Hamiltonian at $\tau =0$, are,
with $\tilde\tau_p=\tau_p+\xi_y$,
\begin{equation}
\label{sys}
\begin{aligned}
   U(-\tilde\tau_p)= U_0\equiv\sqrt{\frac{1}{2}\left(1+\frac{\tilde\tau_p}{\sqrt{\tilde\tau_p^2+|\tilde\Delta_p|^2}} \right)    },\\
   V(-\tilde \tau_p)=V_0\equiv e^{-i {\text {Arg}}(\tilde\Delta_p)}\sqrt{\frac{1}{2}\left(1-\frac{\tilde\tau_p}{\sqrt{\tilde\tau_p^2+|\tilde\Delta_p|^2}}  \right)    }.
\end{aligned}
\end{equation}

The above boundary conditions must be imposed on the general solution, which can be written as a linear combination of the two independent solutions (\ref{ind_sol}) with coefficients $A_p$ and $B_p$.
One then gets
\begin{equation*}
	\begin{aligned}
		&A_p=U_0[U^{(1)}(-\tilde \tau_p)]^*+V_0[V^{(1)}(-\tilde \tau_p)]^*\\
		&B_p=V_0\,U^{(1)}(-\tilde \tau_p)-U_0\,V^{(1)}(-\tilde \tau_p).
	\end{aligned}
\end{equation*}


\begin{figure}
	\centering
	\begin{tabular}{@{}p{\linewidth}}
	\subfigimg[width=\linewidth]{(a)}{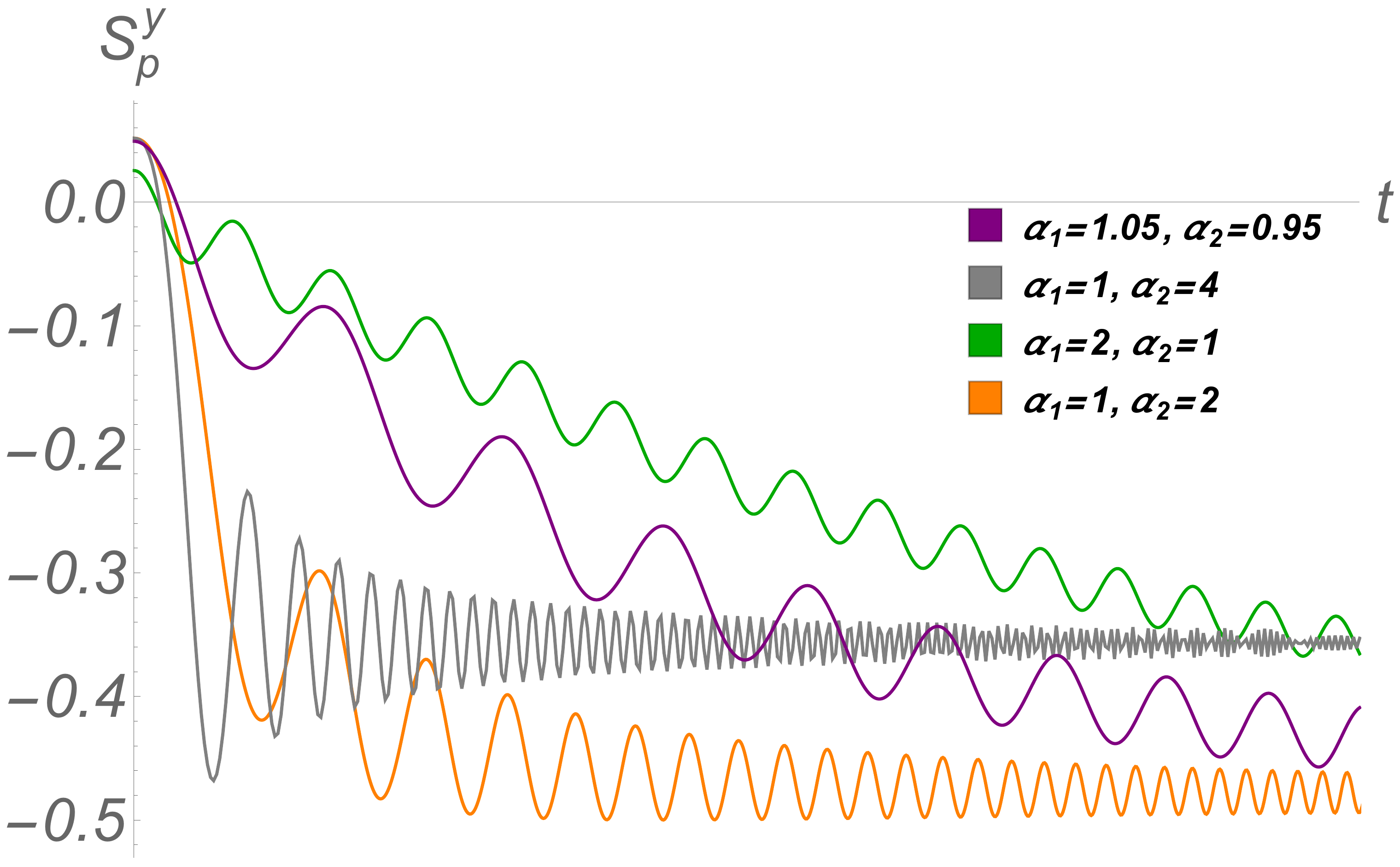} \\
		\vspace{0.25cm}
	\subfigimg[width=\linewidth]{(b)}{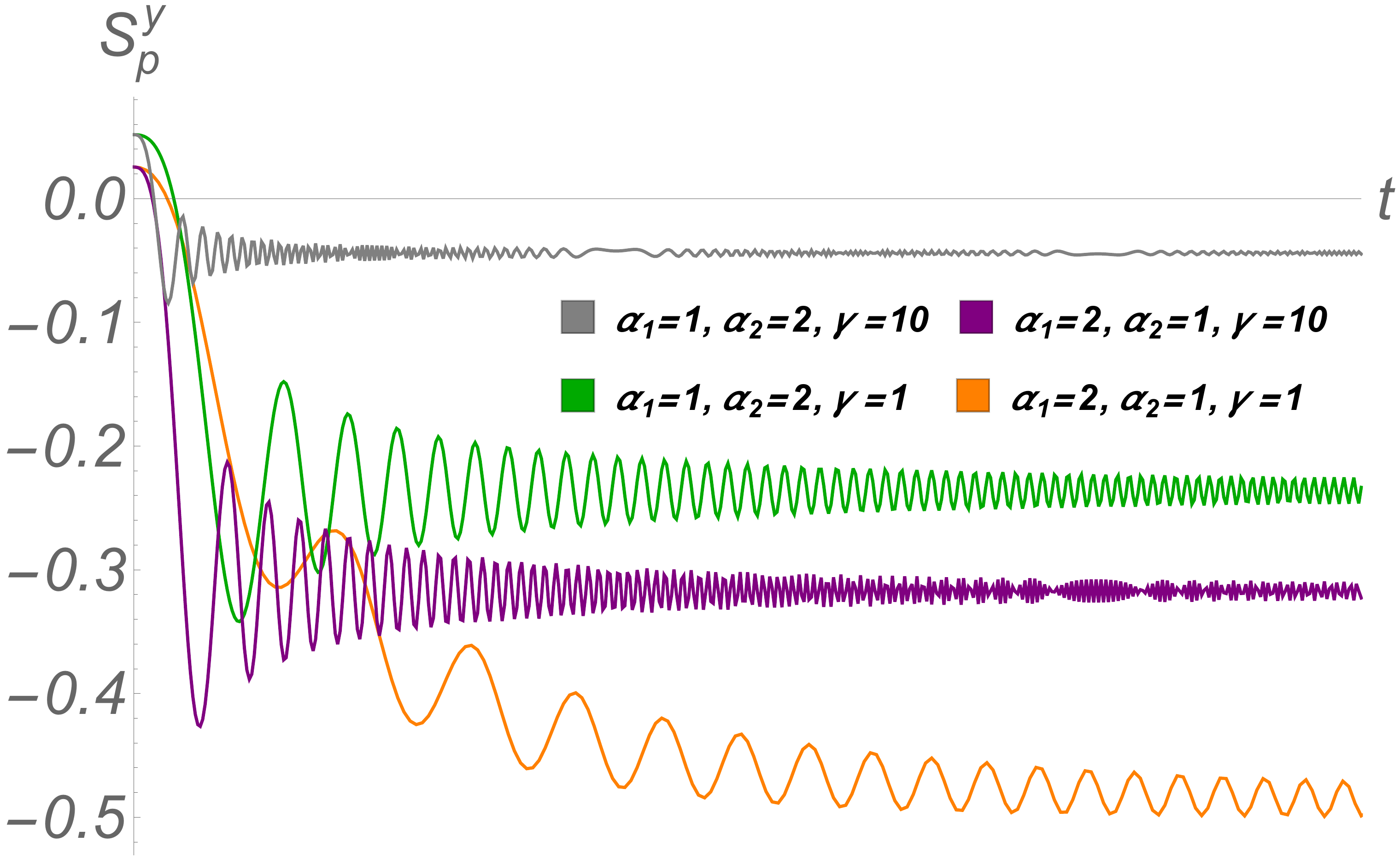}
	\end{tabular}
	\caption{\textbf{Spin polarization component $S^y_{\bf p}$ in different regimes.} Spin $y$-component as a function of time from the solutions of Eq.\ref{SysEq} at fixed momentum and $\theta=\frac{\pi}{2}$. The $z$-component of magnetic field is $B_z=0.3$.  (a) Adiabatic regime with $\gamma_0=0.1$: the Rashba coefficients are $2\alpha_1=\alpha_2$ (orange), $\alpha_1=2\alpha_2$ (green), $0.95\alpha_1=1.05\alpha_2$ (purple) and $4\alpha_1=\alpha_2$ (gray). (b) Quasiadiabatic ($\gamma_0=1$) and nonadiabatic ($\gamma_0=10$): Rashba coefficients are $\alpha_1=2\alpha_2$ (orange, purple) and $2\alpha_1=\alpha_2$ (green, gray).}
	\label{Comp}
\end{figure}

\section{Spin polarization} 
To compute the spin polarization along the \textit{y} direction we use $S^{y}=\Psi^{\dagger}\sigma_z\Psi$ and the analytical solution of the system (\ref{SysEq}), by recalling that the spin quantization axis has been chosen along the $y$ direction~\cite{note2}. Thus, for a given absolute value of the momentum in Eq.(\ref{momentum}), and for $\theta=\frac{\pi}{2}$, the spin along the $y$ direction reads  $S^y_{\bf p}(\tau,p,\theta)=|U(\tau-\tau_p+\xi_z)|^2-\frac{1}{2}$. The numerical computation of the component $S^y_{\bf p}$ as a function of time in different regimes is shown in Fig.\ref{Comp}. 
In the adiabatic regime (Fig.\ref{Comp}a) when $\gamma_0=0.1$ the component along \textit{y} direction of the spin polarization $S^y_{\bf p}$ {\color{black} converges} slowly, with different timescales depending on the relative value of $\alpha_1$ and $\alpha_2$; these also dictate the  frequency of the oscillations. 
\textcolor{black}{The spin adiabatically converges due to the presence of the external electric field.
Initially the spin follows the Rashba field and points in the same direction. With time, the adiabatic influence of the external electric field causes the spin to respond. After a characteristic time, such an influence exceeds that of the Rashba field, thus the spin points along the direction of the electric field. In practice, we observe the saturation of the spin polarization for a given  magnitude of the external electric field.} In the quasiadiabatic or nonadiabatic regimes (Fig.\ref{Comp}b) with $\gamma_0=1$ and $\gamma_0=10$, respectively, convergence occurs more rapidly, with the same qualitative differences due to the distinct coefficients $\alpha_1$ and $\alpha_2$. 
%

\begin{figure*}
	\centering
	\begin{tabular}{@{}p{0.48\linewidth}@{\qquad}p{0.48\linewidth}@{}}
		\subfigimg[width=\linewidth]{(a)}{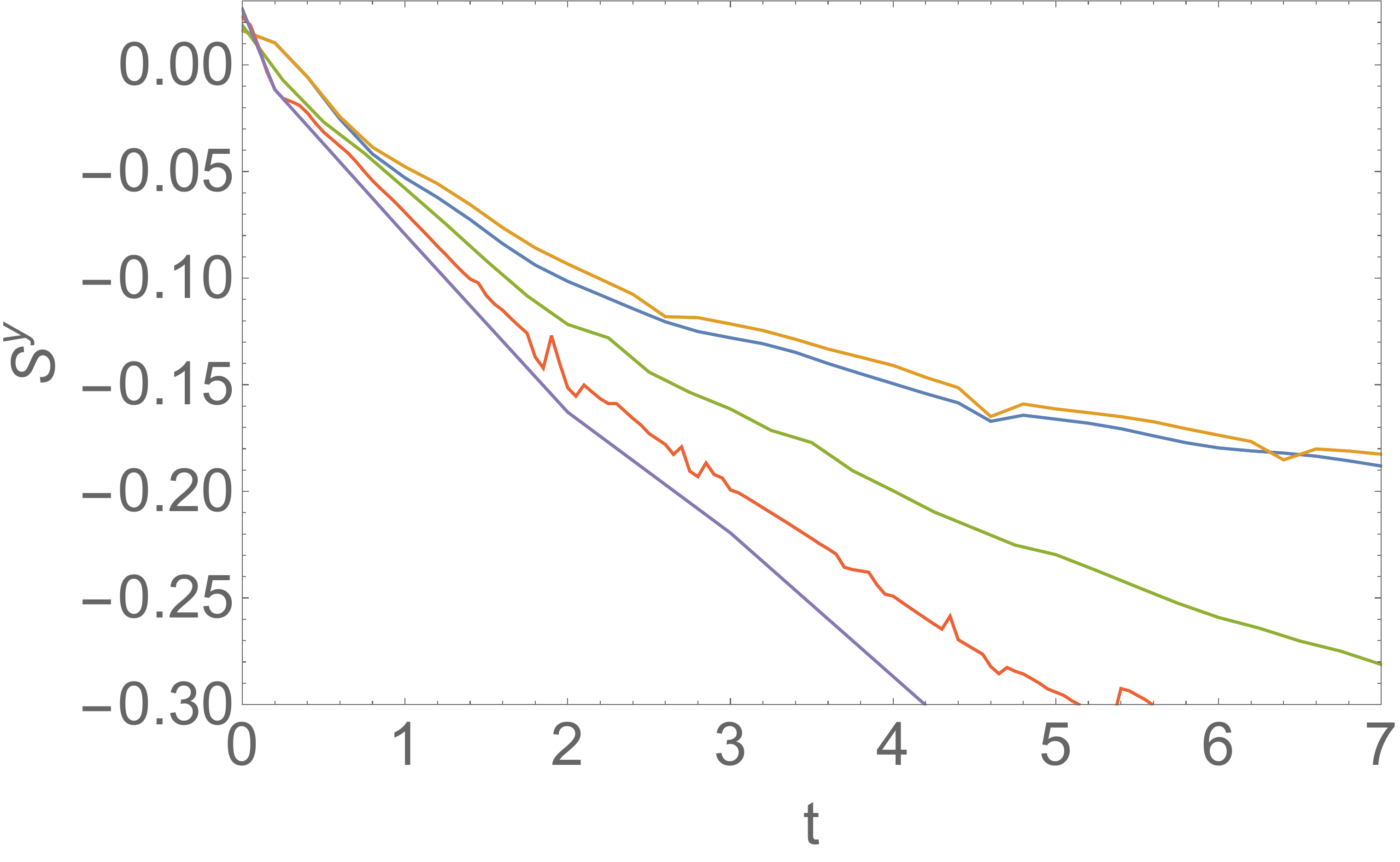} &
		\subfigimg[width=\linewidth]{(b)}{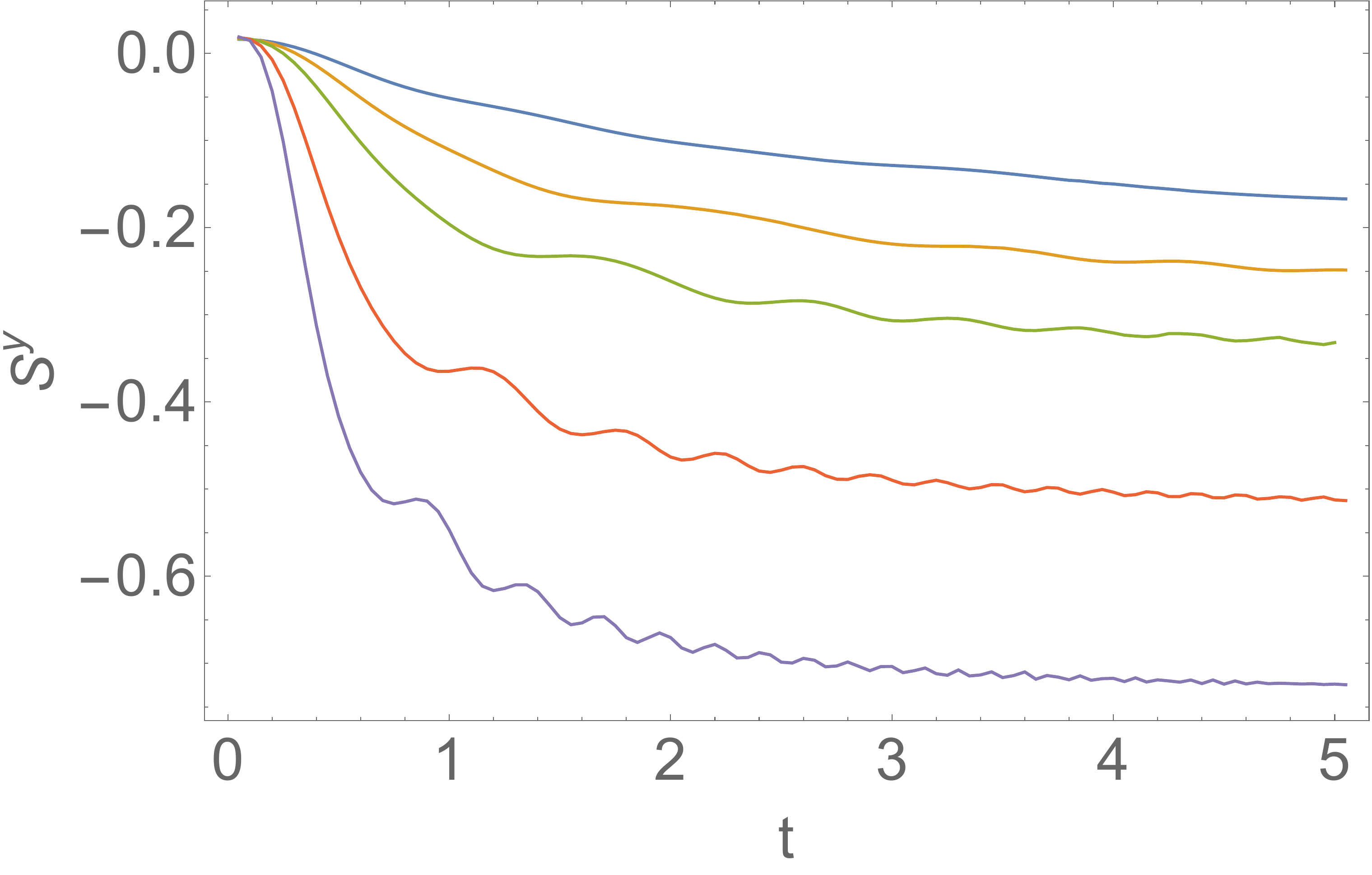}
	\end{tabular}
	\caption{\footnotesize \textbf{Total spin polarization in adiabatic regime $\gamma_0=0.1$.} Total spin polarization  $S^y(t)$ as a function of time $t$ for cold atomic system. (a) Rashba coefficients  $(\alpha_1,\,\alpha_2)$, with $\alpha_1>\alpha_2$, from the upper to the lower curve are blue (1.00001, 0.99999), orange (1.05, 0.95), green (2, 1), red (3, 1) and purple (4, 1). (b) Rashba coefficients $(\alpha_1,\,\alpha_2)$, with $\alpha_1<\alpha_2$, from the upper to the lower curve are blue (0.99, 1.01), orange (1, 1.5), green (1, 2), red (1, 3) and purple (1, 4).}
	\label{Total}%
\end{figure*}

\begin{figure}[bpht!]
	\includegraphics[width=8.5cm, height=5.2cm]{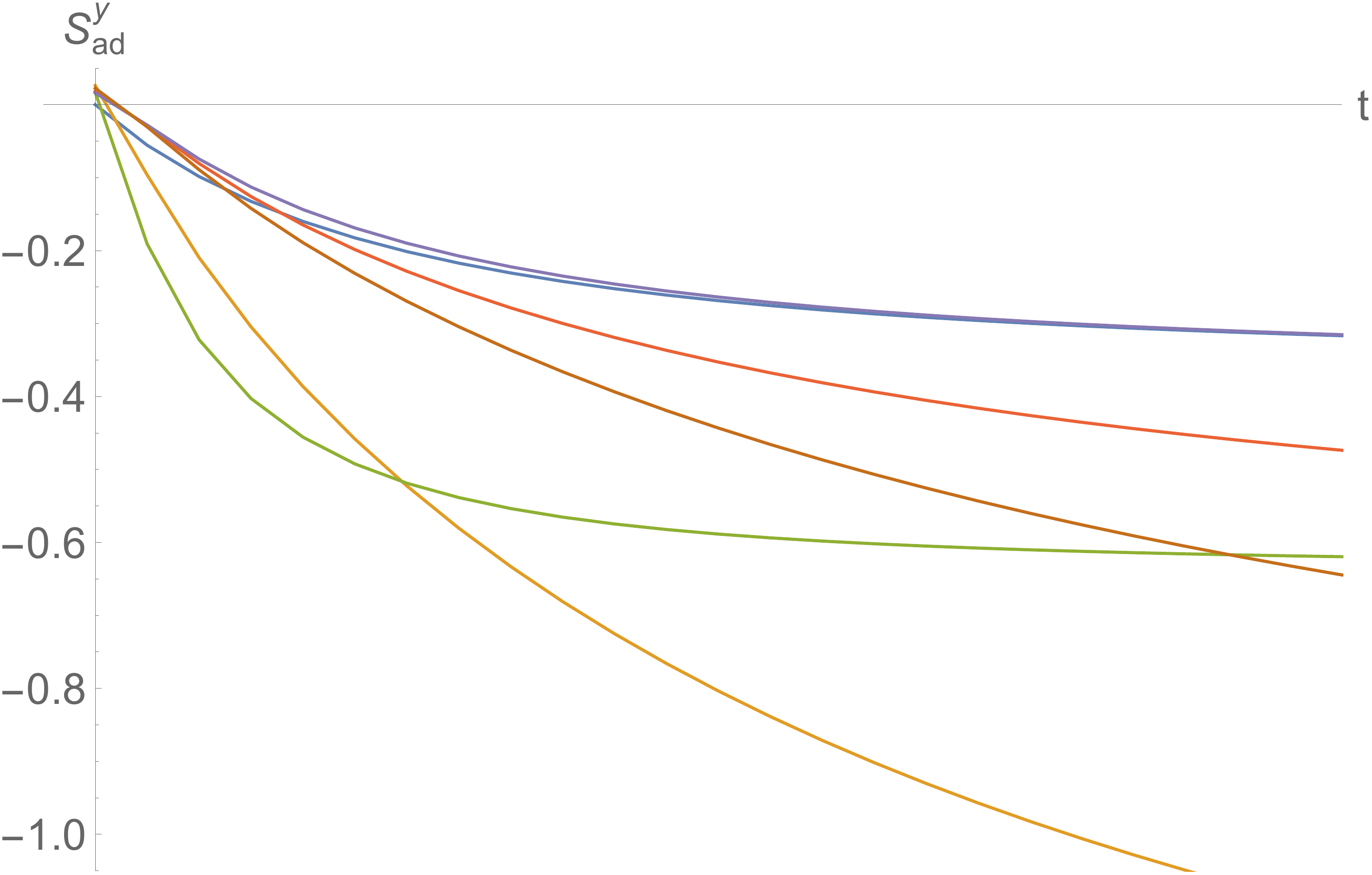}\\
	\caption{\footnotesize \textbf{Adiabatic approximation of the spin polarization.}  $S^y_{ad}$ as a function of time with fixed adiabatic parameter $\gamma_0=0.1$ and different ratios of Rashba coefficients $k=\frac{\alpha_1}{\alpha_2}$ with components of magnetic field {\bf B} = (0.1, 0.2, 0.3) for the purple ($k=0.98$), red ($k=2$), orange ($k=3$), green ($k=0.5$) and yellow ($k=4$) curves. The blue curve is taken at {\bf B} = 0 and $\alpha_1=\alpha_2$, which corresponds to the characteristic result for 2DEG.}
	\label{Adiabatic}
\end{figure}

We now turn our attention to the total spin polarization. Differently from Vignale and Tokatly~\cite{vignale2016theory}, who kept the momentum from Eq.~(\ref{momentum}) fixed  at the Fermi level, assuming that the difference between $p_1$ and $p_2$ is small, we do not fix the momentum in our computation, i.e. the anisotropic case that we consider has the additional variable $p$. We consider the integration within the asymmetric region limited by the energies $E_1(p(\theta,p_F))$ and $E_2(p(\theta,p_F))$, and the Fermi level inthe absence of SOC lays in this region. Thus, the total spin polarization must be integrated as

\begin{equation}
	S^y (\tau)=\sum_{s=1,2} \int \ \frac{d^2p}{(2\pi)^2} \ (-1)^{s-1} \ S^y_{\bf p}(\tau) \Theta \left(\frac{p_F^2}{2m}-E_s({\bf p}) \right)
	\label{Integr}
\end{equation} 
where $\Theta \left(\frac{p_F^2}{2m}-E_s({\bf p}) \right)$ is the Heaviside step function. This means that the total spin polarization can be computed by integrating $S^y_{\bf p}$ over all the angles and averaging over the momentum $p(\theta,p_0)$ as
	\begin{equation}
		S^y ( \tau)= \int_{0}^{2\pi} \ \frac{d\theta}{2\pi}\int_{p_2(\theta,p_0)}^{p_1(\theta,p_0)} \ \frac{p\,dp}{2\pi} \ S^y_{\bf p}(\tau,p,\theta)
		\label{Integr1}
	\end{equation}


Fig.\ref{Total} shows the total spin polarization $S^y$ for the two ratios of Rashba coefficients $\alpha_1>\alpha_2$ (Fig.\ref{Total}a) and $\alpha_1<\alpha_2$ (Fig.\ref{Total}b) in the adiabatic regime $\gamma_0=0.1$. To evaluate Eq.(\ref{Integr}) we consider $m=1$, $p_F=2$ and {\bf B}=(0.1, 0.2, 0.3).
The blue and orange curves in Fig.\ref{Total}a, corresponding to the case $\alpha_1\simeq\alpha_2$, reach a steady condition soon, meaning that the spin direction follows that induced by the electric field. Besides, it confirms the theoretical prediction of the spin polarization for 2DEG with Rashba spin-orbit and Edelstein effect \cite{vignale2016theory}. The violet and red curves in Fig.\ref{Total}a are the total spin polarizations for $\alpha_1=2\alpha_2$, $\alpha_1=3\alpha_2$, and $\alpha_1=4\alpha_2$: they respond very slowly to the influence of an external electric field $E$, hence the steady state of the atomic spin takes longer to build up, depending on the ratio of Rashba coefficients. From Fig.\ref{Total}b we observe that all the curves respond faster than in the case $\alpha_1>\alpha_2$, since saturation occurs in shorter times: only the blue ($1.01\alpha_1=0.99\alpha_2$), orange ($1.5\alpha_1=\alpha_2$) and green ($2\alpha_1=\alpha_2$) curves have a longer build-up time.


Now let us consider the adiabatic approximation and compare it with the solutions in the adiabatic regime of weak electric field $E$. Note that we calculate the spin polarization $S^y_{{\bf p},ad}$ with regard to the lower eigenstate of the system, which means that for all times the system under adiabatical process of the external electric field remains in the instantaneous lower eigenstate. This approach allows to obtain the result of the evolution considering time $\tau$ as a parameter in the calculations. To this purpose we solve the system of equations (\ref{SysEq}) taking into account only the eigenvalue for the lowest state. The component of the approximated spin polarization in the adiabatic regime can be found as
	
\begin{equation}
S^y_{{\bf p},ad}=\frac{1}{2}\frac{|\tilde\Delta_p|^2-\left(\tau-\tilde\tau_p+\sqrt{|\tilde\Delta_p|^2+\tilde\tau_p^2}\right)^2}{|\tilde\Delta_p|^2+\left(\tau-\tilde\tau_p+\sqrt{|\tilde\Delta_p|^2+\tilde\tau_p^2}\right)^2}
\end{equation}

Fig.\ref{Adiabatic} shows the evolution of total spin polarization for a cold atomic system together with that in 2DEG (blue) with {\bf B} = 0 and $\alpha_1=\alpha_2$ with $k=\frac{\alpha_1}{\alpha_2}$, retrieved by integrating numerically over the angles and averaging over the momentum $p$ as in Eq.(\ref{Integr1}) with the substitution of $S^y_{\bf p}(\tau,p,\theta)$ with its adiabatic expression $S^y_{{\bf p},ad}$.

As we can see from Fig.\ref{Adiabatic} the blue curve in the long times coincides with the purple one ($k=0.98$), but in the small times the presence of \textcolor{black}{Rashba SOC field}
makes the latter lay higher than the one for the 2DEG. The green curve ($k=0.5$) saturates quite fast with the small times and even slightly faster than the one for 2DEG, which means that the Edelstein field is more significant than Rashba field. The red ($k=2$), orange ($k=3$) and yellow curve ($k=4$) decline in the slower way and their saturation happens much later in comparison to the purple and green curve, showing that Rashba field still prevails. 


The results obtained from the adiabatic approximation qualitatively coincide with the results calculated from the exact solution, and have the identical subsiding behavior in both small and large ratio of SOC parameters $k$. 
The artificial pseudospin states in cold atoms with Rashba spin-orbit coupling show a longer lifetime, highlighting their better tunability compared with 2DEG.


\section{Discussion} In contrast to solid-state matter, where Rashba spin-orbit coupling depends on the SOC of the material, in cold atoms this type of coupling can be generated synthetically and manipulated externally. The application of artificial gauge fields in quantum gases with neutral atoms provides a series of advantages: lack of disorder compared to 2DEG, convenience for studying many-body systems, external tunability and longer lifetime of pseudospin states. Moreover, a synthetic magnetic field cannot be influenced by a real field due to the absence of dynamical degrees of freedom. In addition, the gauge fields allow to observe a non-equilibrium spin dynamics in many-particle interactions without complications due to the disorder, as we have for 2DEG or other systems in condensed matter systems. 

We described analytically the model of cold atoms in the presence of RSOC and DSOC, \textcolor{black}{weak electric field and external magnetic field represented as a Zeeman field.} After finding the solutions in terms of parabolic cylinder functions and the component of spin polarization, we integrate over all the angles and average over the momentum. Results show that when the ratio of SOC coefficients is $\alpha_1>\alpha_2$ the total spin polarization relaxes much more slowly than for the inverse ratio. Besides, the adiabatic approximation shows that the total spin polarization $S^y_{ad}$ is tunable much better and has longer lifetime of the atomic spin state than the spin state in 2DEG.

While our scheme does not suffer from the heating problem, it may have a problem with the limited lifetime of the pseudospin states, due to atomic collisions which induce a decay of the degenerate dark states into those of lower energy. This difficulty can be addressed, however, by introducing external magnetic fields and weak electric fields. Our results can enable further investigations on spin current, spin Hall effect in cold atoms or other effects requiring stable spin states and large tunability. 

{\bf Acknowledgements}  We thank G. Vignale, C. Tassi, and A. Sheikhabadi for discussion.


\bigskip

\bibliography{Bibliography}

\begin{thebibliography}{10}
\bibitem{note1}
Notice that, for {D}{S}{O}{C} in such solid-state systems, the x and y axes
  correspond the crystal directions (1,1,0) and (-1,1,0), respectively.
\bibitem{note2}
This can obtained from the original {H}amiltonian via the replacement
  $\sigma_x\rightarrow -\sigma_x$, $\sigma_y\rightarrow \sigma_z$,
  $\sigma_z\rightarrow \sigma_y$.
\end{thebibliography}
\bibliographystyle{unsrt}

\end{document}